 \documentclass[twocolumn,prl]{revtex4}
 \usepackage{epsfig}
 \usepackage{graphicx}
 \usepackage{dcolumn}
 \usepackage{amsfonts}
 \usepackage{amsmath}
 \usepackage{amssymb}
 \usepackage{color}
 \usepackage{mathrsfs}
  \newcommand{\beq}{\begin{equation}}
 \newcommand{\eeq}{\end{equation}}
 \newcommand{\ba}{\begin{array}}
 \newcommand{\ea}{\end{array}}
 \newcommand{\beqa}{\begin{eqnarray}}
 \newcommand{\eeqa}{\end{eqnarray}}
 \newcommand{\bal}{\begin{align}}

 \def\Rb{\mathbb{R}}

\usepackage{docs}%
\usepackage{bm}%
\expandafter\ifx\csname package@font\endcsname\relax\else
 \expandafter\expandafter
 \expandafter\usepackage
 \expandafter\expandafter
 \expandafter{\csname package@font\endcsname}%
\fi

\begin{document}

\title{The many-nucleon unified model  and the dynamics of nuclear rotations} 
\author{David J.~Rowe}
\affiliation{Department of Physics, University of Toronto, Toronto, ON M5S 1A7, Canada}

\date{April 25, 2019}

\begin{abstract}
It is determined that a many-nucleon version of the Bohr-Mottelson unified model that contains the essential observables of that model and has irreducible representations that span the Hilbert space of  fully anti-symmetric states of nuclei, is given uniquely by the  symplectic model.  
	This model is shown to  provide a framework for an examination of the dynamics of nuclear rotations.
	A first discovery is that rotational energies are mixtures of potential and  kinetic energies even in an adiabatic limit.
\end{abstract}

 \maketitle

Nuclei have shell structures, similar to  atoms, and rotational states with properties between those of  molecules and superfluids.
	Such properties and the numerous bands of rotational states observed in nuclear spectra  \cite{Morinaga56, BrownG66, HeydeW11, WoodH16eds} 
motivate attempts to understand the dynamics of nuclear rotations and to extend shell-model methods for their description to heavy nuclei, for which the huge dimensions of realistic spherical shell-model spaces currently pose formidable problems.

A fundamental objective is to determine if the rotational energies of nuclei are purely kinetic or mixtures of kinetic and potential energies.
	It is generally assumed  that, like translations, they are kinetic but perturbed by inertial Coriolis and centrifugal forces.
	In the Bohr-Mottelson unified model \cite{BohrM53c, BohrM75}, in its standard adiabatic limit, the states of a  rotational band are assigned a common intrinsic state and, hence, common potential energies.
	A rotating nucleus has also been modelled \cite{RosensteelG02, GraberR03}  as a rotating  Riemann ellipsoid \cite{Chandrasekhar69} with rotational energies given by the kinetic energies of linear combinations of rigid-body and irrotational flows.
	In contrast, two model calculations \cite{ParkCVRR84, BahriR00} showed
the kinetic and potential energy components of rotational energies  to be of comparable magnitude.	
	However, because they used schematic interactions in truncated spaces, their results can be questioned.
	Recall that, as a result of being restricted to single spherical harmonic-oscillator shells, SU(3) model states have zero kinetic-energies and an effective interaction is required to obtain rotational spectra.whereas the rotations of a rigid rotor are 100\% kinetic.	
			
In an extension of the unified model, Nilsson replaced its intrinsic state with an axially symmetric  independent-particle state \cite{Nilsson55}.
	 Such an intrinsic state is viewed in HF (Hartree-Fock) theory as
one of a set of states generated by rotating it through all angles.
	An orthonormal basis of rotational states is then obtained by angular-momentum projection methods 
\cite{Lowdin64, Kamlah68,  LeeC72, HaraS95, BenderH08, Sun16} and
rotational states emerge with mixed kinetic and potential energies.
	However, a many-nucleon version of the unified model based on HF theory
is not ideal, most significantly because its dynamical group of one-body unitary transformations has a single irrep on the whole Hilbert space of a nucleus, whereas the unified model has many representations.
	Ideally, what is needed is a dynamical group that decomposes a nuclear Hilbert space into a sum of subspaces with the property that all matrix elements of essential rotor-model observables  between the states of its different subspaces are zero.	
	Essential observables include the nuclear monopole and quadrupole moments, the angular-momentum operators, the nuclear  kinetic energy, and  SU(2)$_S$ spin and SU(2)$_T$ isospin observables.
 	
	A remarkable fact is that these essential observables are elements of the
already known  Lie algebra  \cite{RosensteelR77a, RosensteelR80, Rowe85}.
of the direct product group		
${\rm Sp}(3,\Rb) \times{\rm SU(2)}_S\times{\rm SU(2)}_T$, where, for an $A$-nucleon nucleus, the Sp$(3,\Rb)$  Lie algebra  is spanned by the  operators, with $1\leq i,j \leq 3$,
\bal 
&
 \hat Q_{ij} = \sum_{n=1}^A \hat x_{ni} \hat x_{nj} , \quad
\hat P_{ij} =  \sum_{n=1}^A ( \hat x_{ni} \hat p_{nj} + \hat p_{ni} \hat x_{nj}),
\label{eq:8.QPL}  \\
&
\hbar \hat L_{ij} 
=  \sum_{n=1}^A \big(\hat x_{ni}\hat p_{nj}  -\hat x_{nj}\hat p_{ni}\big), \quad
\hat K_{ij}= \sum_{n=1}^A  \hat p_{ni}  \hat p_{nj} . \label{eq:8.K} 
\end{align}
	
	As required, a model with these observables has an infinite number of irreps each of which contains a complete set of states of a unified model irrep
with the  property that there are no isoscalar E2 transitions or any non-zero matrix elements of any model observable between states of its different irreps.
	Physical states are inevitably mixtures of states of different model  irreps.
	But, a many-nucleon unified model, defined in this way, has the property that the probabilities, given by the squared matrix elements of an element of its Lie algebra between states of its mixed irreps are the averages of the contributions from the mixed irreps; they cannot be enhanced above these averages by any coherent mixing of the  irreps.

An Sp$(3,\Rb)$ irrep of an $A$-nucleon nucleus is spanned by the positive (or negative) parity eigenstates of a three-dimensional harmonic-oscillator Hamiltonian
 \beq \label{eq:THO}  \textstyle
\hat{\mathcal{H}}(\omega) 
= \tfrac12 \sum_{n=1}^A \sum^3_{i=1} \hbar\omega_i
(c^\dag_{ni} c_{ni} + c_{ni} c^\dag_{ni}) ,
\eeq
where $c^\dag_{ni}$ and $c_{ni}$ are harmonic oscillator raising and lowering operators for which $[c_{ni}, c^\dag_{mj}] = \delta_{m,n} \delta_{i,j}$.
	Then, with corresponding Sp$(3.\Rb)$ operators defined by 
\begin{eqnarray} 
\begin{split}
\textstyle
\hat{\mathcal{A}}_{ij} =  \sum_{n=1}^A c^\dag_{ni} c^\dag_{nj}, \hspace{1cm}
\hat{\mathcal{B}}_{ij} =  \sum_{n=1}^A c_{ni} c_{nj}, \\
\textstyle \hat{\mathcal{C}}_{ij}=\sum_{n=1}^A (c^\dag_{ni} c_{nj}+\tfrac12\delta_{i,j}) ,
 \hspace{1.0cm}
 \end{split}
\end{eqnarray}
for $1\leq  i, j \leq 3$, it follows that a lowest-weight state for an 	Sp$(3.\Rb)$ irrep is defined by the equations
\bal 
\hat{\mathcal{B}}_{ij}  |\sigma,\omega\rangle =0, \quad \:\:
                              & \quad  1\leq i,j \leq 3 ,&  \label{eq:Bvac} \\
\hat{\mathcal{C}}_{ij}  |\sigma,\omega\rangle = 0,\quad \:\: 
                               & \quad  1\leq i < j \leq 3 ,& \label{eq:C1vac} \\
\hat{\mathcal{C}}_{ii}  |\sigma,\omega\rangle = \sigma_i |\sigma\rangle ,
                               & \quad  1\leq i \leq 3 .& 
\label{eq:C2vac}
\end{align}

It is now observed that the unified model is obtained as a mean-field approximation to an  Sp$(3,\Rb)$ irrep in which the values of $\{ \omega_i\}$ are chosen such that the energy $\langle \sigma,\omega| \hat H | \sigma,\omega\rangle$, is minimised for a rotationally invariant many-nucleon Hamiltonian $\hat H$.
	The subset of $\hat{\mathcal{C}}_{ij}$ operators that are angular-momentum operators and do not annihilate the lowest-weight state then act on it to generate a set of rotated states as in standard HF theory and the SU(3) model.
	The $\hat{\mathcal{A}}_{ij}$ raising operators likewise act to generate purely vibrational excitations.
	Thus, by choosing a minimum energy lowest-weight, the first order coupling between the rotational states and their vibrations, including that of the Coriolis and centrifugal forces, is eliminated as assumed in the unified model.
	But, of course,  it is restored in the full symplectic model.
	 
Recall that in mean-field theory, the minimum-energy lowest-weight state  satisfies a self-consistency relationship in which the density of the lowest-energy state has essentially the same shape as that of the mean field of the single-particle Hamiltonian of which it is an eigenstate.
	This relationship was expressed by Bohr, Mottelson  
\cite{BohrM55, BohrM75} in a harmonic-oscillator approximation to the mean field by the equiation	 
\beq \label{eq:shape.consistency}
\sigma_1\omega_1 = \sigma_2\omega_2 = \sigma_3\omega_3 ,
 \eeq
in an interpretation of the Inglis cranking model \cite{Inglis54}.
The remarkable result is that, with its restriction to an Sp$(3,\Rb)$ irrep,
the mean field is identically  that of a harmonic oscillator and equation
(\ref{eq:shape.consistency}) becomes precise.
	 
	The strength of the interaction component of $\hat H$ should also be such that the volume of the shape-consistent state is that expected for near-incompressible nuclear matter.
	Thus, the minimum-energy lowest-weight state is essentially the same for any acceptable choice of the nuclear Hamiltonian.
	 
\begin{figure}[ht]
\centerline{\includegraphics[width=3.5 in]{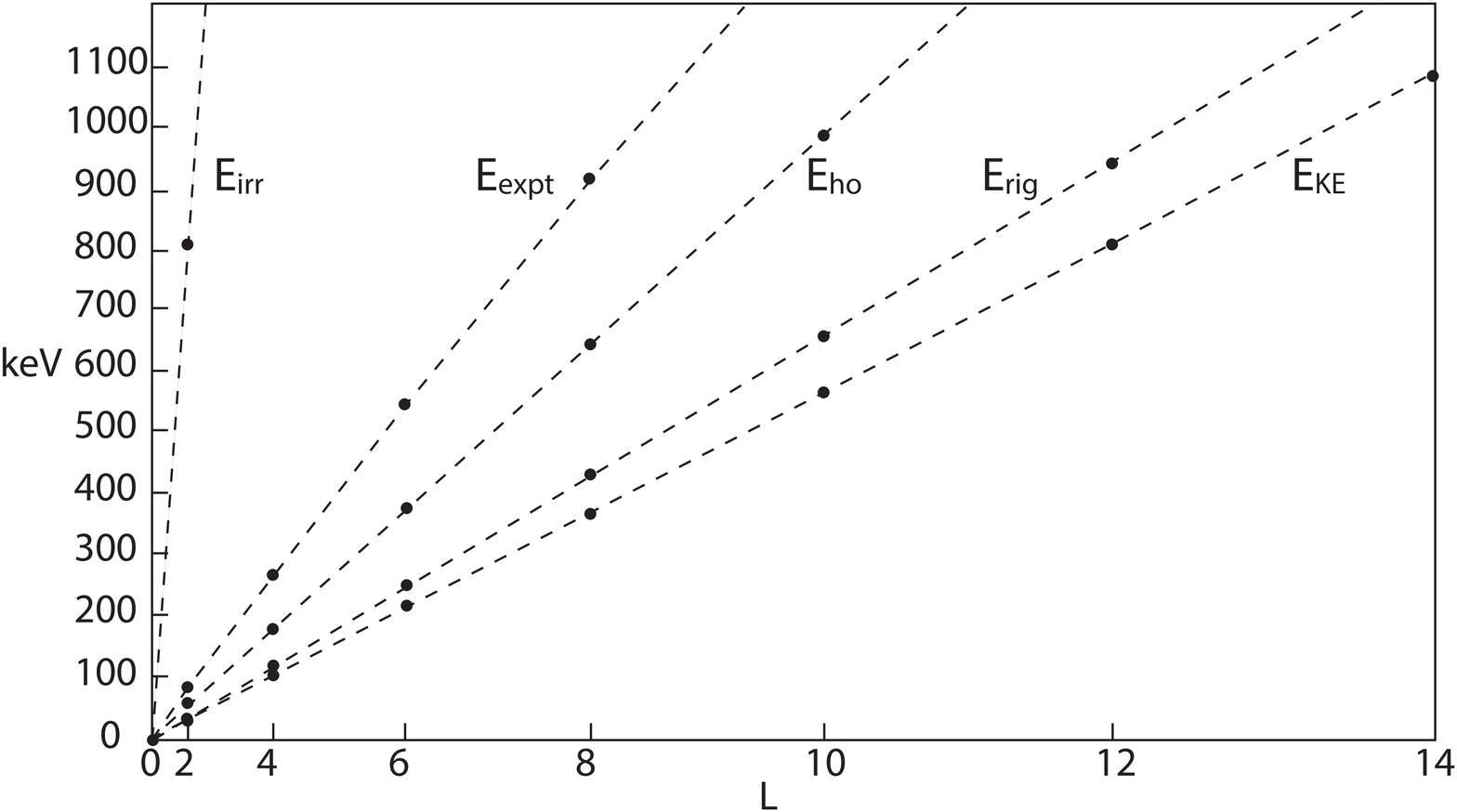}}
\caption{ \label{fig1} 
Excitation energy levels,  ${\rm E}_{\rm expt}$, of the ground-state rotational band of $^{166}$Er for angular-momentum values $0 \leq L \leq 14$ and corresponding energies E$_{\rm irr}$ and E$_{\rm rig}$ for irrotational and rigid-body moaments of inertia for the ${\rm Sp}(3,\Rb)$ irrep 
$\langle 327\tfrac12, 249\tfrac12,249\tfrac12\rangle$.
	The kinetic energies, ${\rm E}_{\rm KE}$, are for states angular-momentum projected from the shape-consistent lowest-weight state for this irrep.
	The  ${\rm E}_{\rm ho}$ energies are those which, in addition to the kinetic energies include the harmonic-oscillator potential energies of the angular-momentum-projected states. }
\end{figure}

To illustrate the possibilities, the precise kinetic energy contributions to the rotational energies of a symplectic model have been calculated for an axially symmetric ${\rm Sp}(3,\Rb) \times{\rm SU(2)}_S  \times {\rm SU(2)}_T$ irrep of spin $S=0$ and minimum isospin.
	This  irrep was previously used \cite{BahriR00} to fit the lower-energy  rotational states of  $^{166}$Er and, in particular, the E2 transitions between them were calculated without the use of an effective charge.
	The chosen irrep is undoubtedly not the most appropriate \cite{JarrioWR91}.
	However, the results obtained are characteristic of those of an axially symmetric irrep that fits the observed quadrupole moments and E2 transition rates.
	The results shown in figure \ref{fig1} 
were derived by use of explicit algebraic expressions for the angular-momentum states projected from an axially symmetric harmonic-oscillator ground state \cite{RoweBB00}.
	Because the many-nucleon kinetic-energy operator is an element of the Sp$(3,\Rb)$ Lie algebra, it was then possible to calculate the kinetic energies of each of the projected angular-momentum states algebraically.
	A remarkable result is that the kinetic energies derived are proportional to $L(L+1)$ to a high level of accuracy.in spite of the fact that the kinetic energy component accounted for only $\sim 35\%$ of the observed rotational energies. This is consistent with the results obtained by Bahri \cite{BahriR00} but is now obtained without use of a truncated space or a schematic interaction.
	
The above results, call for further and more detailed investigation which are now much simplified by the use of a mean-field basis for an Sp$(3,\Rb)$ irrep.
	Desirable calculations would be to compare the results of full many-nucleon calculations in spaces of  single symplectic-model irreps with those of corresponding mean field approximations with the same Hamiltonian.
	Initial calculations might be for the axially symmetric rotational states based on the Hoyle state of $^{12}$C, the first excited state of $^{16}$O or the ground state of $^{20}$Ne with realistic interactions.	
	Extensions to the more general triaxial irreps and irreps with spin-orbit interactions will be more challenging.
	Hopefully, with the major advances, in which it has become possible to identify the states of an Sp$(3,\Rb)$ irrep in SA-NCSM (symmetry-adapted no-core shell model) calculations \cite{DytrychLDRWRBLB18}, such calculations, and others for triaxial irreps will soon be feasible.
	A very satisfying many-nucleon theory of deformed nuclei will then have been achieved.
	
\acknowledgements	

\bibliographystyle{apsrev}
\bibliography{master}

\end{document}